\documentclass[prl,aps,twocolumn]{revtex4}

\usepackage{pxfonts}
\usepackage{graphicx}
\begin{document}

\title{Excitation Energy Dependence of the Exciton Inner Ring}

\author{Y.\,Y. Kuznetsova, J.\,R. Leonard, L.\,V. Butov}
\affiliation{Department of Physics, University of California at San
Diego, La Jolla, California 92093-0319, USA}
\author{J. Wilkes, E.\,A. Muljarov}
\affiliation{Department of Physics and Astronomy, Cardiff University, Cardiff CF24 3AA, United Kingdom}
\author{K.\,L. Campman, A.\,C. Gossard}
\affiliation{Materials Department, University of California at Santa Barbara, Santa Barbara, California 93106-5050, USA}

\date{\today}

\begin{abstract}
\noindent
We report on the excitation energy dependence of the inner ring in the exciton emission pattern. The contrast of the inner ring is found to decrease with lowering excitation energy. Excitation by light tuned to the direct exciton resonance is found to effectively suppress excitation-induced heating of indirect excitons and facilitate the realization of a cold and dense exciton gas. The excitation energy dependence of the inner ring is explained in terms of exciton transport and cooling.
\end{abstract}


\date{\today}

\maketitle

\section{I. Introduction}
An indirect exciton in a coupled quantum well structure (CQW) is a bound pair of an electron and a hole confined to spatially separated QWs. The small overlap of the electron and the hole wavefunctions increases the lifetime of indirect excitons by orders of magnitude compared to the lifetime of regular direct excitons. In addition, indirect excitons are oriented dipoles with a built-in dipole moment $ed$, where $d$ is close to the separation between the QW centers. The repulsive dipole-dipole interaction between indirect excitons allows them to screen in-plane disorder in the structure \cite{Ivanov02}. As a result, indirect excitons can travel over large distances before optically decaying~\cite{Hagn95, Butov98, Larionov00, Butov02, Voros05, Ivanov06, Gartner06, Gartner07, High07, High08, Vogele09, Hammack09, Remeika09, Grosso09, High09prl, Kuznetsova10, Alloing11, Winbow11, Alloing12}. The long lifetimes of indirect excitons also allow them to cool down to low temperatures, well below the temperature of quantum degeneracy $T_{\rm dB} = (2 \pi \hbar^2 n_{\rm X})/(M_{\rm X} g k_{\rm B})$~\cite{Butov01} (in the studied CQW with the exciton mass $M_{\rm X} = 0.22m_0$ and the spin degeneracy $g=4$, $T_{\rm dB} \sim 3$K for the density per spin state $n_{\rm X}/g = 10^{10}$ cm$^{-2}$). Furthermore, the built-in dipole moment of an indirect exciton allows control of exciton transport by voltage \cite{Hagn95, Gartner06, High07, High08, Remeika09, Grosso09, High09prl, Winbow11} and light \cite{Kuznetsova10}. The long lifetime, large transport distances, efficient cooling, and control of transport make indirect excitons a model system for studying transport of cold bosons in condensed matter materials.

Studies of indirect excitons have resulted in observations of patterns in the exciton emission, including the inner ring~\cite{Butov02, Ivanov06, Stern08, Hammack09, Alloing12}, external ring~\cite{Butov02, Butov04, Rapaport04, Chen05, Haque06, Yang10}, localized bright spots~\cite{Butov02, Butov04, Yang10}, and macroscopically ordered exciton state~\cite{Butov02, Butov04, Levitov05, Yang06} (ring shapes are specific to point excitation geometry). The inner ring has been explained in terms of exciton transport and cooling: laser excitation heats the exciton gas, excitons cool towards the lattice temperature as they travel away from the excitation spot, the cooling results in an increase in the occupation of the low-energy optically active exciton states \cite{Feldmann87, Hanamura88, Andreani90} and, in turn, the appearance of an emission ring around the excitation spot~\cite{Butov02, Ivanov06, Hammack09}. In this paper, we report on the study of the excitation energy dependence of the inner ring in the exciton emission pattern. We find that by adjusting the excitation energy to the direct exciton resonance, we can reduce heating due to laser excitation, facilitating the realization of a cold and dense exciton gas.

\section{II. Experimental data}
The CQW structure used in this experiment contains two 8 nm GaAs QWs separated by a 4 nm Al$_{0.33}$Ga$_{0.67}$As barrier. The CQW diagram is given in Fig.~1a. The sample was grown by molecular-beam epitaxy. The effective spacing between the electron and hole layers is given by $d=11.5$ nm~\cite{Butov99}, close to the nominal distance between the QW centers. Details on the CQW structure can be found in Ref.~\cite{Butov99}.

Indirect excitons were photogenerated by a tunable Ti:Sapphire laser focused to a 2.8~$\mu$m spot on the sample. Excitation energy $E_{\rm ex}$ was varied between 1.55 and 1.85~eV. For this excitation below the Al$_{0.33}$Ga$_{0.67}$As barrier energy, no excess free carriers are photogenerated, and both the external ring and the localized bright spots are absent~\cite{Butov04, Rapaport04, Chen05, Haque06, Yang10}. Exciton density was controlled by the laser excitation power $P_{\rm ex}$. Spatial resolution was 1.4~$\mu$m. The spectra were measured with a liquid nitrogen cooled charge-coupled device (CCD) placed after a double-grating spectrometer. The measurements were performed at an applied gate voltage of 1.2~V, at which the indirect excitons are approximately 20 meV lower in energy than direct excitons in the structure. The cryostat bath temperature was 1.6~K.

Figure 1b shows an intensity profile $I(x)$ of the inner ring in the emission of indirect excitons. The major characteristics of the inner ring are indicated: $R$ is the half width at half maximum (HWHM), $I_{\rm max}$ is the average maximum intensity, and $\Delta I$ is the depth of the inner ring. The contrast $\Delta I/I_{\rm max}$ of the inner ring gives a measure of heating of the exciton gas by laser excitation as detailed below.

Figure 1c shows $x$-energy images of the exciton emission for varying $P_{\rm ex}$. A higher emission energy corresponds to a higher exciton density because of the repulsive dipole-dipole interaction between indirect excitons \cite{Ivanov02}. For the considered densities, the emission linewidth is $\lesssim 2$ meV. It corresponds to the exciton emission linewidth, which is determined by the homogeneous and inhomogeneous broadening and can be small \cite{High09nl}. In comparison, the linewidth characteristic for the emission of an electron-hole plasma is close to the sum of the electron and hole Fermi energies in the plasma and is considerably larger \cite{Butov91}. The narrow emission line, which is characteristic for the inner ring reported in Refs.~\cite{Butov02, Ivanov06, Stern08, Hammack09} and in the present paper, indicates that in all these experiments the inner ring forms in a system of excitons rather than in electron-hole plasma~\cite{Hammack09, Ivanov2010}.

\begin{figure}
\begin{center}
\includegraphics[width=8.5cm]{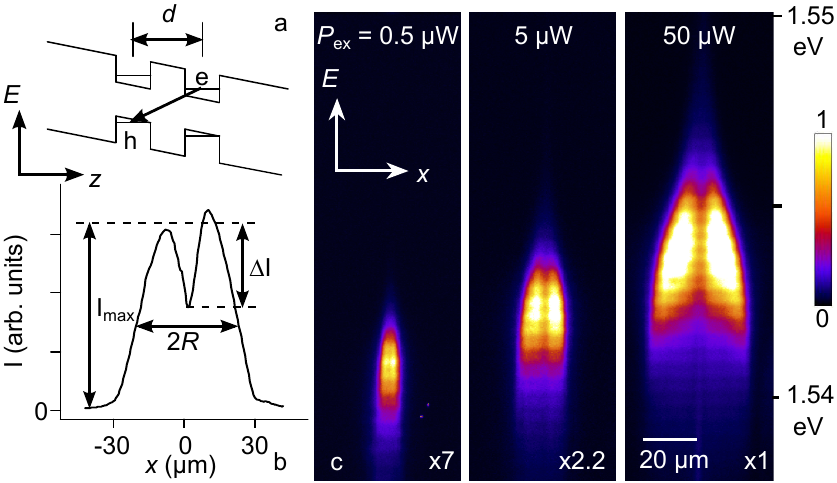}
\caption{(a) CQW band diagram; e, electron; h, hole. (b) Profile of the inner ring in the emission of indirect excitons; $R$ is the half width at half maximum (HWHM); $I_{\rm max}$ is the average maximum intensity; $\Delta I$ is the depth of the ring. (c) $x$-energy images of the exciton emission for different excitation powers $P_{\rm ex}$. Excitation energy $E_{\rm ex}=1.588$ eV. The position of the laser excitation spot is $x$ = 0.}
\end{center}
\end{figure}

Figure 2a presents the spatially integrated emission intensity of indirect excitons $I_{\rm total}$ as a function of excitation energy $E_{\rm ex}$.  $I_{\rm total}(E_{\rm ex})$ follows the absorption curve of the CQW, with resonances at the excitation energy equal to the heavy hole direct exciton (hh DX) energy $E_{\rm ex} = 1.570$ eV and light hole direct exciton (lh DX) energy $E_{\rm ex} = 1.588$ eV. Figures 2b-f present the excitation power dependence of the inner ring at these values of $E_{\rm ex}$ along with a higher nonresonant excitation energy $E_{\rm ex} = 1.676$ eV. Both the ring extension $R$ (Fig. 2b) and contrast $\Delta I/I_{\rm max}$ (Fig. 2c) increase with increasing excitation density $P_{\rm ex}$. This behavior is understood as follows: At low exciton densities, indirect excitons are localized by the in-plane disorder potential in the structure and do not travel away from the excitation spot. The size of the exciton cloud is close to the size of the excitation spot. No inner ring is observed in this regime (Figs. 2d-f, blue curves). At high densities, excitons screen the disorder potential. Delocalized excitons travel away from the excitation spot and cool toward the lattice temperature during their travel. This results in the enhancement of the occupation of low-energy optically active exciton states \cite{Feldmann87, Hanamura88, Andreani90} and, in turn, the appearance of the ring in the exciton emission pattern \cite{Butov02, Ivanov06, Hammack09} (Figs. 2d-f, orange and black curves).

\begin{figure}
\begin{center}
\includegraphics[width=8.5cm]{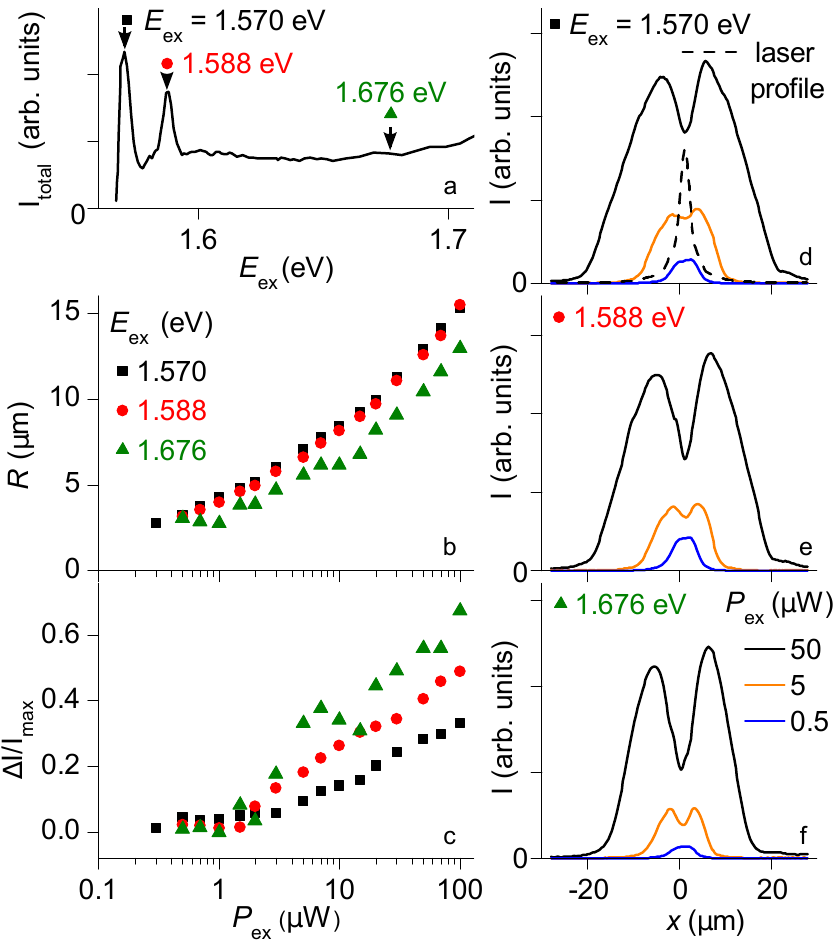}
\caption{(a) Spatially integrated emission intensity of indirect excitons as a function of excitation energy $E_{\rm ex}$ for $P_{\rm ex}$ = 50 $\mu$W. (b) HWHM $R$ and (c) contrast $\Delta I/I_{\rm max}$ of the inner ring as a function of excitation power $P_{\rm ex}$ for $E_{\rm ex} = $ 1.570 eV (black $\square$), 1.588 eV (red $\circ$) and 1.676 eV (green $\triangle$), indicated in (a). (d-f) $P_{\rm ex}$ dependence of the indirect exciton emission profile for the above values of $E_{\rm ex}$. The laser excitation profile is shown by the dashed line in (d).}
\end{center}
\end{figure}

The detailed excitation energy dependence of the integrated emission intensity $I_{\rm total}$, ring extension $R$, and contrast $\Delta I/I_{\rm max}$ is presented in Figure~3. Increased absorption at the hh DX and lh DX resonances leads to a higher exciton population at $E_{\rm ex}=1.570$ and 1.588~eV, conveyed by the peaks in $I_{\rm total}(E_{\rm ex})$ (Fig. 3a). A higher exciton population leads to a broader emission pattern and corresponding peaks in $R(E_{\rm ex})$ (Fig. 3b). Figure~3c shows the ring contrast $\Delta I/I_{\rm max}$ as a function of excitation energy. A strong reduction of the ring contrast is observed at low $E_{\rm ex}$, indicating a strong reduction of the excitation-induced heating of the exciton gas. The highest exciton density (Fig. 3a), the largest ring extension (Fig. 3b), and the smallest ring contrast (Fig. 3c) are observed for the excitation tuned to the hh DX resonance. At $E_{\rm ex}$ below the hh DX resonance, the inner ring vanishes as a result of reduced absorption (Fig. 3a-c).

\begin{figure}
\begin{center}
\includegraphics[width=8.5cm]{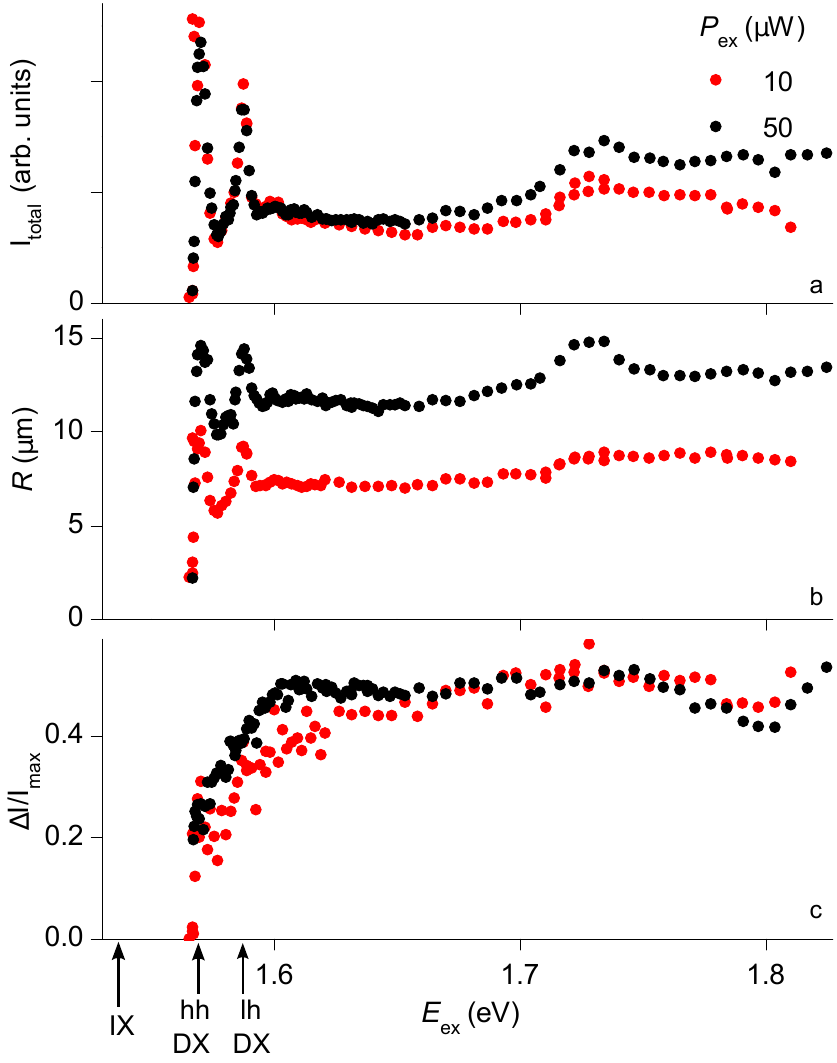}
\caption{(a) Integrated emission intensity $I_{\rm total}$, (b) extension $R$, and (c) contrast $\Delta I/I_{\rm max}$ of the inner ring as a function of excitation energy $E_{ex}$ for excitation power $P_{\rm ex} = 10~\mu$W (red) and $50~\mu$W (black). The data for 10 and $50~\mu$W in (a) are normalized at $E_{\rm ex} = 1.6$ eV.}
\end{center}
\end{figure}

\section{III. Numerical simulations}
Following the approach developed in \cite{Ivanov02, Ivanov06, Hammack09}, we solved a set of coupled equations to find the spatial profiles of the indirect exciton density, temperature and emission intensity. In the steady-state scenario relevant to the experiment, the indirect exciton density $n_X$ satisfies the following transport equation:
\begin{equation}
\nabla \left[ D \nabla n_{\rm X} + \mu n_{\rm X} \nabla (u_0 n_{\rm X}) \right] - \Gamma_{\rm opt} n_{\rm X} + \Lambda = 0.
\label{eq1}
\end{equation}
The two terms in the square brackets of equation (\ref{eq1}) are due to the diffusion and drift currents of excitons, respectively. The disorder-dependent diffusion coefficient $D = D_{\rm 0} {\rm exp}[-U_{\rm 0}/(u_{\rm 0} n_{\rm X} + k_{\rm B} T_{\rm X})]$ accounts for the ability of dipole-orientated indirect excitons to screen the disorder potential. Here, $u_{\rm 0} n_{\rm X}$ is the dipole-dipole interaction potential and $U_{\rm 0} = 1.4$~meV is the amplitude of the disorder potential. The exciton mobility is given by the generalized Einstein relation, $\mu = D (e^{T_{\rm dB}/T_{\rm X}} - 1)/(k_{\rm B} T_{\rm dB})$ \cite{Ivanov02}. The optical decay rate $\Gamma_{\rm opt}(T_{\rm dB},T_{\rm X})$ incorporates the fact that only excitons in the lowest energy states may decay optically \cite{Feldmann87, Hanamura88, Andreani90}. The generation rate of excitons, $\Lambda(r)$ follows the Gaussian profile of the laser where $r$ is the distance from the center of the excitation spot. The transport equation (\ref{eq1}) is solved in 2D geometry with radial symmetry of the density assumed.

\begin{figure}
\begin{center}
\includegraphics[width=8.5cm]{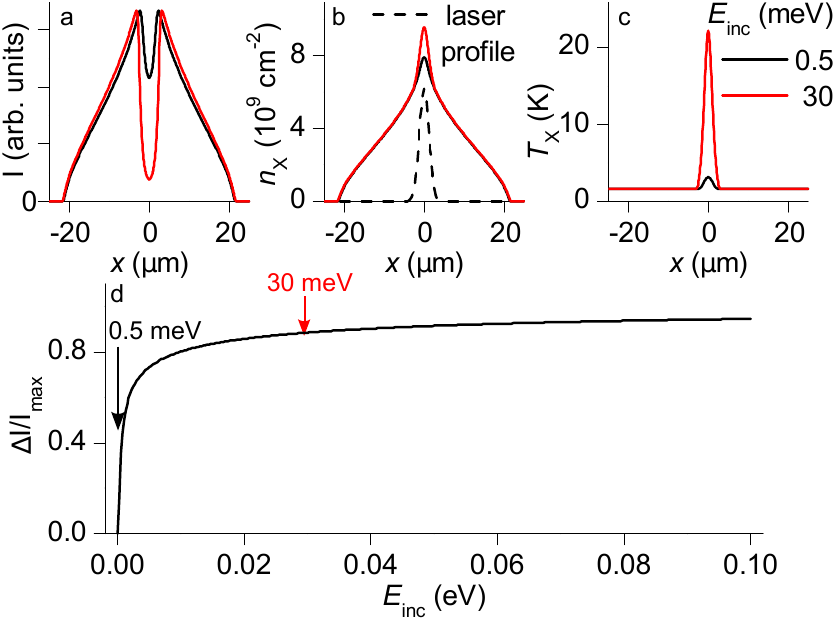}
\caption{Theoretical simulations for photoluminescence intensity (a), density (b), and temperature (c) of indirect excitons as a function of $x$ for $E_{\rm inc} = E_{\rm ex} - E_{\rm IX}$ = 0.5~meV (black) and 30~meV (red). (d) Contrast of the inner ring $\Delta I/I_{\rm max}$ as a function of $E_{\rm inc}$ calculated from the simulated exciton emission profiles.}
\end{center}
\end{figure}

The exciton temperature $T_{\rm X}$, which stands in the expressions for the diffusion coefficient, mobility, and decay rate is found by solving the following heat balance equation,
\begin{equation}
S_{\rm phonon}(T_{\rm dB},T_{\rm X}) = S_{\rm pump}(T_{\rm dB},T_{\rm X},\Lambda,E_{\rm inc}).
\label{eq2}
\end{equation}
Here, $S_{\rm phonon}$ is the rate of cooling of indirect excitons by a bath of bulk longitudinal acoustic phonons, which act to restore the exciton temperature to the lattice temperature. Within the excitation spot, heating occurs due to the injection of high energy excitons, which then thermalize within the timescales much shorter than the cooling time associated with acoustic phonons \cite{Ivanov99}. The heating rate $S_{\rm pump}$ is characterized by the pump rate $\Lambda$ and the excess energy acquired by photoexcited excitons $E_{\rm inc}$, which increases with photon energy. Expressions for $S_{\rm phonon}$, $S_{\rm pump}$ and $\Gamma_{\rm opt}$ and all other parameters are given in \cite{Hammack09}. In previous studies, the heating/cooling rate which arises from optical decay of indirect excitons was found to be small in comparison to $S_{\rm phonon}$ and $S_{\rm pump}$ \cite{Hammack09}. It is therefore neglected in these calculations.

From the numerical solution to equations (\ref{eq1},\ref{eq2}), one can extract the emission intensity $I$, density $n_{\rm X}$, and temperature $T_{\rm X}$ of indirect excitons, shown as functions of $x$ in Figs. 4a, 4b, and 4c, respectively, for two different excess energies $E_{\rm inc}$. The contrast of the inner ring obtained from these calculations for various values of $E_{\rm inc}$ is presented in Fig.~4d.

The model illustrates the sensitivity of the exciton temperature to the tuning of the laser. Exciton cooling via scattering with acoustic phonons becomes less effective against the heating supplied by excitation with increasing excitation energy.

\section{IV. Discussion}
Since the intensity reduction in the center of the inner ring is due to a higher exciton temperature in the region of laser excitation, the contrast of the inner ring $\Delta I/I_{\rm max}$ gives a measure of the exciton heating. Figure 3 shows that tuning the laser excitation to the hh DX resonance $E_{\rm ex}$ = 1.570~eV both maximizes the light absorption and minimizes the laser-induced heating of excitons, thus facilitating the realization of a cold and dense exciton gas.

There is a qualitative agreement between the observed and simulated shapes of emission profiles (see Fig. 2d-f and Fig. 4a). The simulated ring contrast $\Delta I/I_{\rm max}$ increases with excess energy $E_{inc}$ (Fig. 4d), and therefore with excitation energy, due to increasing exciton temperature in the excitation spot (Fig. 4c). An increase of the ring contrast with excitation energy is in qualitative agreement with experimental data (see Fig. 3c and Fig. 4d). Note however that the simulations are based on a simplified qualitative model, which includes only exciton relaxation via emission and absorption of acoustic phonons and only one energy band - the band of hh indirect excitons. Such model qualitatively explains the main observed characteristics of the inner ring, including the enhancement of the ring contrast with excitation energy, reported in this paper. A quantitative comparison with the experiment requires a theory which takes into account exciton relaxation via emission and absorption of optical phonons as well as hh and lh bands of direct and indirect excitons. The development of such theory forms the subject for a future work.

\section{V. Summary}

We studied the excitation energy dependence of the inner ring in the indirect exciton emission pattern. Such ring forms due to exciton transport and cooling. We found that the excitation-induced heating of the exciton gas is suppressed at lower excitation energies, reaching a minimum at the excitation energy tuned to the heavy hole direct exciton resonance, below which the inner ring disappears due to the diminished light absorption in the structure. We found that tuning the excitation laser to the heavy hole direct exciton energy both increases the density of indirect excitons and effectively suppresses the laser-induced heating of indirect excitons, thus facilitating the realization of a cold and dense exciton gas.

\section{Acknowledgements}
We thank Nikolay Gippius, Aaron Hammack, and Leonidas Mouchliadis for discussions. Support of this work by NSF and EPSRC is gratefully acknowledged.

\end{document}